\documentclass[12pt]{article}
\usepackage{color}
\usepackage[dvips]{graphicx,psfrag}
\usepackage{amsmath,amssymb}
\usepackage{bm}

\baselineskip=\normalbaselineskip
\renewcommand{\baselinestretch}{1.34}
\makeatletter
\@addtoreset{equation}{section}
\makeatother

\renewcommand{\thefootnote}{\fnsymbol{footnote}}

\newcommand{\ket}[1]{\bigl|#1\bigr>}
\newcommand{\bra}[1]{\bigl<#1\bigr|}

\newcommand{\maru}[1]
{{\ooalign{\hfil#1\/\hfil\crcr\raise.167ex\hbox{\mathhexbox20D}}}}

\newcommand{\cD}{{\mathcal{D}}}

\newcommand{\cH}{{\mathcal{H}}}
\newcommand{\cI}{{\mathcal{I}}}
\newcommand{\cJ}{{\mathcal{J}}}

\newcommand{\cL}{{\mathcal{L}}}

\newcommand{\cN}{{\mathcal{N}}}
\newcommand{\cO}{{\mathcal{O}}}

\newcommand{\cS}{{\mathcal{S}}}

\newcommand{\cV}{{\mathcal{V}}}

\newcommand{\bCC}{{\mathbb{C}}}

\newcommand{\bC}{{\boldsymbol{C}}}

\newcommand{\bH}{{\boldsymbol{H}}}

\newcommand{\bJ}{{\boldsymbol{J}}}

\newcommand{\bP}{{\boldsymbol{P}}}

\newcommand{\bZ}{{\boldsymbol{Z}}}

\newcommand{\bx}{{\boldsymbol{x}}}
\newcommand{\by}{{\boldsymbol{y}}}

\newcommand{\bunit}{{\boldsymbol{1}}}
\newcommand{\bzero}{{\boldsymbol{0}}}

\newcommand{\del}{\partial}

\newcommand{\eq}[1]{(\ref{#1})}

\newcommand{\nn}{\nonumber}

\DeclareMathOperator{\tr}{tr}
\DeclareMathOperator{\Tr}{Tr}

\allowdisplaybreaks[1]

\newcommand{\nol}
  {
    \begin{array}{r}
    \raisebox{-1.5mm}{\mbox{\scriptsize{$\centerdot$}}} \\ 
    \raisebox{3.5mm}{\mbox{\scriptsize{$\centerdot$}}}
    \end{array}
   \!\!}
\newcommand{\nor}
  {\!\!
    \begin{array}{l}
    \raisebox{-1.5mm}{\mbox{\scriptsize{$\centerdot$}}} \\ 
    \raisebox{3.5mm}{\mbox{\scriptsize{$\centerdot$}}}
    \end{array}
   }

\newcommand{\ad}{{\rm ad}}
\newcommand{\Ad}{{\rm Ad}}
\newcommand{\Del}[1]{{\frac{\delta}{\delta {#1}}}}
\newcommand{\V}{{V}}
\newcommand{\J}{{\cJ}}

\makeatletter
\@addtoreset{equation}{section}
\@addtoreset{theorem}{section}
\@addtoreset{definition}{section}
\@addtoreset{lemma}{section}
\@addtoreset{proposition}{section}
\def\@seccntformat#1{\csname the#1\endcsname.\quad}
\makeatother

\begin{document}
\begin{flushright}
\parbox{35mm}{%
KUNS-2104 \\
{\tt arXiv:0711.4191}
}
\end{flushright}

\vfill

\begin{center}
{\large{\bf 
Notes on the Hamiltonian formulation 
of 3D Yang-Mills theory \\
}}
\end{center}

\vspace{5mm}

\begin{center}
\large
{{Masafumi Fukuma}}$^1$\footnote%
{E-mail: {\tt fukuma@gauge.scphys.kyoto-u.ac.jp}},~\,%
{{Ken-Ichi Katayama}}$^1$\footnote%
{E-mail: {\tt katayama@gauge.scphys.kyoto-u.ac.jp}}~\,and~\,%
{{Takao Suyama}}$^2$\footnote%
{E-mail: {\tt suyama@phya.snu.ac.kr}}
\normalsize
\vspace{7mm}

$^1~$Department of Physics, Kyoto University, 
Kyoto 606-8502, Japan 

\vspace{2mm}
$^2~$Department of Physics, 
Seoul National University, Seoul 151-747, Korea

\end{center}
\vfill
\renewcommand{\thefootnote}{\arabic{footnote}}
\setcounter{footnote}{0}
\addtocounter{page}{1}

\begin{center}
{\bf Abstract}
\end{center}

\begin{quote}

Three-dimensional  Yang-Mills theory is investigated 
in the Hamiltonian formalism based on the Karabali-Nair variable. 
A new algorithm is developed 
to obtain the renormalized Hamiltonian 
by identifying local counterterms in Lagrangian
with the use of fictitious holomorphic symmetry 
existing in the framework with the KN variable. 
Our algorithm is totally algebraic 
and enables one to calculate the ground state wave functional 
recursively in gauge potentials. 
In particular, the Gaussian part thus calculated is shown 
to coincide with that obtained by Leigh et al. 
Higher-order corrections to the Gaussian part are also discussed. 
\end{quote}
\vfill
\renewcommand{\baselinestretch}{1.4}
\newpage

\section{Introduction}

Understanding nonperturbative aspects of Yang-Mills theory 
is one of the most important themes in modern theoretical physics. 
It has been playing the role of theoretical arena 
where many new ideas are tested. 
Also, in the attempts there have been obtained new analytical tools, 
each of which in turn always has opened a way 
to vast area of new applications.

In the present paper, we investigate 3D Yang-Mills theory. 
This is certainly a  simplified toy model 
for a realistic 4D theory, 
but still shares many of essential features with 4D YM theory, 
including infrared slavery, asymptotic freedom, color confinement 
and the existence of mass gap. 
Furthermore, the 3D theory has physically meaningful degrees of 
freedom, as opposed to more simplified 2D YM theory.

Recently there have been remarkable developments 
in the understanding of 3D YM theory \cite{LMY}\cite{Gr}. 
Among them is the study \cite{LMY} based on the Hamiltonian formalism 
with the so-called Karabali-Nair (KN) variable \cite{KKN}\cite{AKN}. 
(For the analysis based on the Kogut-Susskind Hamiltonian 
see, e.g., \cite{3D}.)

The KN variable is a sort of Bars' corner variable \cite{Ba}, 
which is local and gauge-invariant. 
In their original papers \cite{KKN}, 
Karabali, Kim and Nair investigated the vacuum wave functional 
of 3D pure $SU(N)$ YM theory, 
and determined its behaviors at the IR and UV regimes. 
They obtained the string tensions for finite $N$'s, 
which are in good agreement with lattice results \cite{LTW}. 
Recently, with a few Ans\"{a}tze 
Leigh, Minic and Yelnikov \cite{LMY} 
determined completely the Gaussian part of the vacuum wave functional 
in the 't Hooft limit, 
and obtained glueball mass spectrum. 
Their arguments are somehow intriguing,  
but the obtained spectrum exhibits excellent agreement with lattice data \cite{LTW}.

As will be reviewed briefly in Section 2,  
the most subtle part in the discussions of Leigh et al.\ 
is the renormalization of the Schr\"{o}dinger equation,  
to remove divergences 
which always arise when two successive local variations 
act on the wave functional. 
This is usually carried out term by term 
on all possible terms in the wave functional, 
as was done in \cite{LMY}.

In the present paper, 
we propose a new algorithm to obtain the renormalized 
Schr\"{o}dinger equation, 
which is certainly effective in the 't Hooft limit. 
We heavily use the fictitious holomorphic symmetry \cite{KKN}\cite{LMY}
existing in the framework based on the KN variable, 
and determine the possible local counterterms 
for the Lagrangian rewritten in terms of the KN variable. 
Then, in the 't Hooft limit the Schro\"dinger equation 
simply becomes the Hamilton-Jacobi equation 
for the resulting effective Lagrangian.

The equation thus obtained actually has a different form  
from the original one of Karabali and Nair, 
but for the Gaussian part of the ground state wave functional 
it gives the same result with that of Leigh et al. 
We believe that this gives a justification for their result. 
Since our equation is totally algebraic 
and easy to solve recursively in gauge potentials, 
one can easily determine higher-order corrections 
to the Gaussian part, as will be demonstrated in section \ref{ground_state}. 
Inclusion of other fields and application to deconstructing 4D YM theory 
will be reported in our subsequent papers.

The present paper is organized as follows. 
In section 2 we review the framework of Karabali, Kim and Nair \cite{KKN}, 
and stress the importance of local counterterms 
which may appear when the product of variations in the Hamiltonian 
are normal-ordered. 
In section 3 we introduce a new algorithm to calculate 
such local counterterms, 
restarting from the bare Lagrangian 
rewritten in terms of the KN variable. 
We show that the form of local counterterms are almost determined 
by a simple reasoning based on dimensions and holomorphic invariance. 
We expect that this is the simplest method to obtain them 
in the 't Hooft limit. 
We then write down the equation that determines wave functionals. 
In section 4 we show that our equation can be solved recursively 
in gauge potentials to give the ground state wave functional. 
We further show that the resulting Gaussian part 
coincides with that of Leigh et al. 
Higher-order corrections to the Gaussian part are also discussed. 
Section 5 is devoted to conclusions and outlook. 
A few Appendices are added where detailed explanations 
are provided for some formulas used in text.

\section{3D Yang-Mills theory and Karabali-Nair variable}

We consider 3D Yang-Mills theory with gauge group $G=SU(N)$. 
Spacetime coordinates are denoted by 
$x=(x^\mu)=(\bx,t)$ with $\bx=(x^1,x^2)$, 
and gauge potentials by traceless antihermitian matrices 
belonging to ${\rm su}(N)$,
\begin{align}
 A_\mu(x) &= \frac{1}{i} t^a A^a_\mu(x) = -A_\mu^\dag(x),
\\
 F_{\mu\nu}(x) &= \frac{1}{i} t^a F^a_{\mu\nu}(x)
                 =\bigl[ \del_\mu+A_\mu(x),\,\del_\nu+A_\nu(x)\bigr]
               = - F_{\mu\nu}^\dag(x) \nn\\
 & \bigl( F^a_{\mu\nu} = \del_\mu A^a_\nu-\del_\nu A^a_\mu 
     + f^{abc} A^b_\mu A^c_\nu \bigr) .
\end{align}
Here the hermitian matrices $t^a$ $(a=1,\cdots,N^2-1)$ are 
normalized as $\tr t^a t^b = (1/2)\,\delta^{ab}$, 
so that $f^{acd} f^{bcd}=N\,\delta^{ab}$  
for $ \bigl[t^a,\,t^b\bigr]=i f^{abc}\,t^c$. 
Note that for $X=(1/i)\, t^a X^a$ and $Y=(1/i)\, t^a Y^a$, 
$\bigl[ X, Y\bigr]=(1/i)\, t^a \bigl[X,Y\bigr]{}^a$ 
with $\bigl[X,Y\bigr]{}^a=f^{abc} X^b Y^c$,  
and 
$\bigl[X,t^b\bigr]=t^a (\ad X){}^{ab}$ 
with $(\ad X){}^{ab}= f^{acb}X^c$.
We also denote $X\cdot Y\equiv X^a\,Y^a=-2\tr (XY)$. 
We often write $X\cdot X$ as $X^2$.

We take the temporal gauge, $A_0(x)\equiv 0$, 
after which there only exists the residual gauge symmetry, 
$A_i(\bx,t)\to g^{-1}(\bx)\,A_i(\bx,t)\,g(\bx)$ $(i=1,2)$ 
with $g(\bx)\in G=SU(N)$. 
In order to define the Karabali-Nair (KN) variable 
from the remaining gauge potentials $A_i(\bx,t)$, 
we first introduce complex coordinates 
$z\equiv x^1+i x^2$,  $\bar z\equiv x^1 - i x^2$ 
and often write them as $\bx=(z,\bar z)$.  
They give
\begin{align}
    \del \equiv \del_z = \dfrac{1}{2}(\del_1 - i \del_2),\qquad
    \bar \del \equiv \del_{\bar z} = \dfrac{1}{2}(\del_1 + i \del_2),
\end{align}
and accordingly, we combine the gauge potentials as
\begin{align}
     A(x) &\equiv A_z(x) = \frac{1}{2}\bigl(A_1(x) - i A_2(x) \bigr), \\
     \bar A(x) &\equiv A_{\bar z}(x) 
             = \frac{1}{2}\bigl(A_1(x) + i A_2(x) \bigr) = - A^\dag(x).
\end{align}
There always exists a matrix-valued function $\V(x)\in G^\bCC=SL(N,\bCC)$
such that 
\begin{align}
 A = -\del V^\dag\, \V^\dag{}^{-1}, \qquad
 \bar A = \V{}^{-1}\, \bar\del \V 
  ~~~\bigl( = - A^\dag\bigr) .
\label{VbarV}
\end{align}
One can easily show that $\V$ is unique up to the left multiplication 
of a holomorphic matrix:
\begin{align}
 V(z,\bar z, t) \rightarrow h(z,t) \,\, \V(z,\bar z,t) 
  \qquad \bigl( h(z,t)\in SL(N,\bCC) \bigr),
\label{hol_V}
\end{align}
while the right multiplication of a time-independent $SU(N)$ matrix 
corresponds to a residual gauge transformation under the temporal gauge: 
\begin{align}
 &
    \V(z,\bar z, t) \rightarrow \V(z,\bar z,t) \,\, 
    g(z,\bar z)
  \qquad \bigl( g(z,\bar z) \in SU(N) \,\bigr) 
  \nn\\
 & \Bigl(
    A \to g^{-1}A\, g + g^{-1}\del g,\quad
    \bar A \to g^{-1}\bar A\, g + g^{-1}\bar\del g \Bigr).
\end{align}
Theory thus must be invariant under 
\begin{align}
 \V(z,\bar z, t) \rightarrow 
   h(z,t) \,\, \V(z,\bar z,t) \,\, g(z,\bar z)
   \quad
   \bigl( h(z,t)\in SL(N,\bCC),\, g(z,\bar z)\in SU(N)\bigr).
\end{align}  
The KN variable \cite{KKN} is then defined by
\begin{align}
 H(x) = \V(x)\, \V^\dag(x).
\end{align}
This is a {\em gauge-invariant, local} variable, 
and transforms under the holomorphic transformation \eq{hol_V} as
\begin{align}
 H(z,\bar z,t) \to h(z,t)\,H(z,\bar z,t)\, \bar h(\bar z,t)
 \quad
 \bigl( \bar h(\bar z,t)\equiv \bigl( h(z,t) \bigr)^\dag \bigr).
\label{hol_H}
\end{align}
With the KN variable, the gauge potentials are written as 
\begin{align}
 A &= \V{}^{-1}\,\del \V + \V{}^{-1} \J \, \V ,
\label{A}
\\
 \bar A &= \V{}^{-1}\, \bar\del \V ,
\label{A-bar}
\end{align}
where the hermitian connection $\J(x)$ is defined by%
\footnote{
This $\J(x)$ differs from the original hermitian connection $J(x)$ 
introduced by Karabali and Nair \cite{KKN} 
by a multiplicative factor as
$\J(x)=-(\pi/N)\,J(x)$.
}
\begin{align}
 \J \equiv -\del H\, H^{-1}.
\label{J}
\end{align}
This transforms under the holomorphic transformations \eq{hol_V} as 
\begin{align}
 \J(z,\bar z,t) &\to h(z,t)\,\J(z,\bar z,t)\,h^{-1}(z,t)
  + h(z,t)\,\del h^{-1}(z,t).
\label{hol_J}
\end{align}
Equation \eq{VbarV} gives the local change of variables 
from $(A,\bar A)$ to $\bigl(\V, \V^\dag)$, 
while eqs.\ \eq{A}--\eq{J} should be regarded 
as defining the local change of variables 
from $(A,\bar A)$ to $(V,H)$, 
not to $(V,\J)$. 
This is because the latter viewpoint does not reflect 
the relation $\bar A=-A^\dag$.

The Hamiltonian under the temporal gauge, $A_0\equiv 0$, 
is then given formally by 
\begin{align}
 \bH
 &=\int_{\bx}\Bigl[ -\frac{g^2}{2} \frac{\delta}{\delta A^a(\bx)}
  \frac{\delta}{\delta \bar A^a(\bx)}
  -\frac{2}{g^2}(F^a_{\bar z z}(\bx))^2\Bigr].
\end{align}
Using the Gauss law operator
\begin{align}
 \hat{\cI}^a(\bx)\equiv \bigl(\ad D_i\bigr)^{ab}\frac{\delta}{\delta A^b_i(\bx)}
  =\bigl(\ad D_z\bigr)^{ab}\frac{\delta}{\delta A^b(\bx)}
   +\bigl(\ad D_{\bar z}\bigr)^{ab}\frac{\delta}{\delta {\bar A}^b(\bx)},
\end{align}
we have 
$\delta/\delta \bar A= (\ad D_{\bar z})^{-1}\,
\bigl(-\ad D_z \,\delta/\delta A + \hat{\cI}\bigr)$. 
Setting $D\equiv \del + \J$, we have
\begin{align}
 &\ad D_z = \cV^{-1}\,\ad D\,\,\cV, \quad~
  \ad D_{\bar z} = \cV^{-1}\,\ad\bar\del \,\,\cV, 
\\
 &\Del{A} = \cV{}^{-1} \Del{\J},\quad~
  \Del{\bar A} = 
   -\,\cV{}^{-1}(\ad\bar \del)^{-1}\, \ad D\Del{\J} 
   +  (\ad D_{\bar z})^{-1} \,\hat{\cI}, 
\end{align}
where $\cV(\bx)=(\cV^{ab}(\bx))$ represents $\Ad\,V(\bx)$, 
$V(\bx)\,t^b\,V^{-1}(\bx)=t^a\,\cV^{ab}(\bx)$.  
Thus, when acting on gauge-invariant wave functionals 
$(\hat\cI=0)$, 
the Hamiltonian becomes
\begin{align}
 \bH &= \int \Bigl[\, \frac{g^2}{2} \frac{\delta}{\delta \J^a}\cdot
  (\ad\bar\del)^{-1}\, \ad D^{ab}\Del{\J^b}
  -\frac{2}{g^2}(\bar\del \J^a)^2\Bigr].
\label{formal}
\end{align}
This bare Hamiltonian has only a formal sense  
because it always gives rise to a short-distance singularity 
$\bigl(\propto \delta^2(\bzero)\bigr)$ 
when acting on a local functional of $\J(\bx)$, 
and thus needs to be normal-ordered. 
By writing the wave functional as 
\begin{align}
 \Psi[\J(\bx)]=\exp\Bigl[\,\frac{2\pi N}{(g^2 N)^2} P[\J(\bx)]\Bigr],
\end{align}
the Schr\"odinger equation becomes
\begin{align}
 \frac{g^2}{2} E
  = \int\Bigl[\,
   + \frac{1}{4m^2}\,\frac{\delta P}{\delta \J}\cdot
     (\ad \bar\del)^{-1}\,\ad D\,\frac{\delta P}{\delta \J}
   + \frac{\pi}{2N}
      \nol  \Del{\J}\cdot(\ad\bar\del)^{-1}\,\ad D\,
     \frac{\delta P}{\delta \J} \nor 
   - (\bar\del \J)^2
   \Bigr],
\label{naive}
\end{align}
where $m$ is the (dimensionful) 't Hooft coupling constant, 
$m\equiv g^2 N/2\pi\,$.
The second term on the right-hand side is seemingly of order $1/N$ 
but cannot be neglected even in the 't Hooft limit 
($N\to\infty$ with $m$ fixed finite). 
This is because counterterms coming from planar diagrams 
are accompanied by the sum over loop index of order $N$ 
so that the net result can be of order $N^0=(1/N)\times N$. 
Note that such additional terms should be local 
since they stem from short-distance singularities.
Once such local counterterms are extracted,%
\footnote{
Karabali and Nair claimed that there should arise 
an additional term in \eq{naive} 
as $(1/2)\int_\bx \J\cdot \delta P/\delta \J$ \cite{KKN}. 
This agrees with our renormalized Hamiltonian (to be given later) 
in the IR limit. 
Further discussions on this point will be given at the end of section 3. 
}
the second term can be neglected 
safely in the 't Hooft limit.

One can guess the form of the local counterterms. 
We first note that finite local counterterms in the Hamiltonian formalism 
will appear when two successive functional derivatives 
are applied to the same exponent 
(of the form $N\tr\mbox{(something)}$) 
or when the functional derivative acts on a regulator. 
Thus, the finite local counterterms in the Hamiltonian, 
$\bH_{\rm loc}=-(g^2/2)\int \bigl[(\delta/\delta A^a_i)^2\bigr]_{\rm loc}$, 
will take the form 
$\bH_{\rm loc}\sim g^2 N\int 
\bigl[ N f(A,\del_j A) + f_i^a(A,\del_j A)\delta/\delta A^a_i\bigr] $,%
\footnote{
The overall $N$ comes from the loop-index sum, 
and the $N$ in the parenthesis comes from the exponent.
}
and will not include terms proportional to $(\delta/\delta A^a_i)^2$.  
This can be thought of as a correction to the original Lagrangian, 
whose form will be roughly estimated 
(using the relation $E_i=(1/i)\delta/\delta A_i \sim (1/g^2)\dot A_i$ 
although this should be modified after inclusion of time-derivative terms)
as
\begin{align}
 \cL_{\rm loc}\sim g^2 N^2 f(A,\del_j A) 
  + iN \dot A^a_i f^a_i(A,\del_j A)
\label{L_loc_guess}
\end{align}
with no terms proportional to $\dot A_i^2$.
This expression will hold even when the theory is rewritten 
with the KN variable, 
so that we expect that the main part of the $\cL_{\rm loc}$ 
consists of dimension-two terms which does not include time derivatives 
(recall that $g^2 N$ has mass dimension one) 
and dimension-three terms which is linear in the time derivative of a field.

\section{Effective Hamiltonian method}

Usually the renormalization in the Hamiltonian formalism 
is carried out with the following steps:

\begin{enumerate}

\item
Start from a given bare Lagrangian

\item
Obtain the classical Hamiltonian

\item
Solve the Schr\"odinger equation

\end{enumerate}

\noindent
The third step needs renormalizing the wave functional term by term, 
which is actually a cumbersome procedure to carry out.

Since we only need to extract finite local counterterms, 
the following algorithm to obtain the renormalized effective 
Hamiltonian turns out to be possible: 

\begin{enumerate}

\item
Start from a given bare Lagrangian

\item
Find the effective Lagrangian including finite local counterterms

\item
Solve the Hamilton-Jacobi equation for the effective Hamiltonian

\end{enumerate}

\noindent
As opposed to usual quantum field theories in four dimensions,  
the second step is actually manageable  
since 3D gauge theory is super renormalizable; 
local counterterms can be controlled in the 't Hooft limit 
simply by a symmetry argument and dimensional analysis, 
as we see below.

We again start from the Lagrangian
\begin{align}
 \cL_{\rm YM} &= \frac{1}{2g^2} \tr\bigl( F_{\mu\nu}^2 \bigr) 
  = -\frac{1}{4g^2}\bigl(F^a_{\mu\nu}\bigr)^2 
\nn\\
 &= \frac{1}{2g^2}\,\Bigl[
  \bigl(\dot A_i^a+[A_0,A_i]^a\bigr)^2-\bigl( B^a\bigr)^2
  \Bigr]
\end{align}
and consider the partition function
\begin{align}
 Z&=\int\frac{[dA_\mu ]}{\rm Vol(gauge)}\,e^{i S_{\rm YM}[A_\mu]}~~.
\end{align}
We take the temporal gauge $A_0(x)\equiv 0$, 
which leads to 
\begin{align}
 Z&=\int\frac{[dA_\mu ]}{\rm Vol(res.\ gauge)}\,
  {\rm Det}\bigl(\delta A^\omega_0/\delta\omega|_{\omega=0}\bigr)\,
  \delta[A_0]\,e^{i S_{\rm YM}[A_\mu]}
\nn\\
 &=\int\frac{[dA_1 dA_2 ]}{\rm Vol(res.\ gauge)}\,
  {\rm Det}^{1/2}(\del_0^2)\,
  e^{i S_{\rm YM}[A_1,A_2]}
\end{align}
with
\begin{align}
 S_{\rm YM}[A_1,A_2 ]=\int\! d^3x \,\frac{1}{2g^2}\,
  \bigl( \dot A_i^2 - B^2 \bigr).
\end{align}

With the change of variables \eq{A}--\eq{J}, 
we have
\begin{align}
   \delta A & =  \V{}^{-1}\,\bigl( 
    \delta \J + \bigl[ D,\delta V V^{-1}\bigr]\bigr)\, \V
\nn\\
    & =  \V{}^{-1}\,\bigl( 
    \bigl[ D,-\delta H H^{-1}\bigr] + \bigl[ D,\delta V V^{-1}\bigr]\bigr)\, \V,
\label{variation0a}
\\
   \delta\bar A & =  \V{}^{-1}\, \hspace{5mm}
       \bar\del \bigl( \delta V V^{-1}\bigr)\hspace{7.5mm}\, \V, 
\label{variation0b}
\end{align}
where $-\delta V\, V^{-1}$ and $-\delta H\, H^{-1}$ are the Maurer-Cartan form, 
and we denote $\omega_\mu\equiv -\del_\mu V\, V^{-1}$. 
The metric over the functional space then becomes
\begin{align}
 ||\delta A\,\delta\bar A||
  &\equiv -\int_x\tr \delta A\delta\bar A 
\nn\\
 &= \int_x\tr\Bigl[ 
  \bigl(\delta V V^{-1}-(1/2)\Delta^{-1}
\bar\del \ad D \delta H H^{-1}\bigr)\cdot\Delta\,
  \bigl(\delta V V^{-1}-(1/2)\Delta^{-1}\bar\del\ad D \delta H H^{-1}\bigr)
\nn\\
 &~~~~~~~~~~~~~~
   - (1/4)\,\delta H H^{-1}\cdot\ad D\ad\bar\del\,\Delta^{-1}\,
\ad D\ad\bar\del\,\delta H H^{-1}\Bigr],
\end{align}
where $\Delta$ is the Laplacian in the adjoint representation:%
\footnote{
In the following $\ad\,\bar\del$ will often be abbreviated as $\bar\del$.
}
\begin{align}
 \Delta\equiv \frac{1}{2}\bigl(\ad D\,\ad\bar\del + \ad\bar\del\,\ad D\bigr).
\end{align}
The Haar measures of $V$ and $H$ are given by 
$||\delta V||^2\equiv -\int \tr \bigl(\delta V V^{-1}\bigr)^2$ and
$||\delta H||^2\equiv -\int \tr \bigl(\delta H H^{-1}\bigr)^2$, respectively, 
and thus the functional measure becomes%
\footnote{
The measure $[dV dH]$ ($H=V V^\dag$) is a natural extension 
of the area element $dx\wedge dy$ 
written in terms of the complex coordinate $z=x+iy=r\,e^{i\theta}$ and 
its absolute square $r^2=|z|^2$:
\begin{align}
 \int [dz\,d(r^2)]\,\cdots 
  \equiv \frac{i}{2}\,\oint \frac{dz}{z}\int_0^\infty\! d(r^2)\,\cdots
  ~\Bigl(=\int\! dx\wedge dy \,\cdots\Bigr).
\nn
\end{align}
}
\begin{align}
 \bigl[ dA \,d\bar A\bigr]
  =\bigl[dV dH \bigr]\cdot
   {\rm Det}\bigl(-\ad\bar\del\,\ad D\bigr).
\end{align}
We also have
\begin{align}
   \dot A & = \V{}^{-1}\,\bigl( 
    \dot \J - \bigl[ D,\omega_0\bigr]\bigr)\, \V, \\
   \dot{\bar A} & = \V{}^{-1}\, \hspace{5mm}
      \bigl(-  \bar\del \omega_0\bigr)\,\hspace{5.5mm} \V .
\end{align}

Putting everything together, we obtain
\begin{align}
 Z &= 
  \int \frac{[dV dH]}{\rm Vol(hol.\ trsf)}\,
  {\rm Det}^{1/2}(\del_0^2)\cdot
  {\rm Det}\bigl(-\ad\bar\del\,\ad D\bigr)
  \cdot e^{i S_0[\omega_0,\J]}
\label{partition_function}
\end{align}
with 
\begin{align}
 S_0&=\frac{1}{2g^2}\int\!\!d^3x\Bigl[
  - 4\bigl(\dot \J-[D,\omega_0]\bigr)\cdot\bar\del\omega_0 
  + 4\,\bigl(\bar\del \J)^2\Bigr]
\nn\\
 &=\frac{1}{2g^2}\int\!\!d^3x\Bigl[
  - 4\,\omega_0\cdot\Delta\omega_0 
  - 4\,\bar\del\omega_0\cdot\dot \J
  + 4\,\bigl(\bar\del \J)^2\Bigr].
\end{align}
One can easily see that $S_0$ is invariant under 
the time-dependent holomorphic transformation \eq{hol_V}--\eq{hol_J}, 
since $\omega_0$ transforms 
as $\omega_0\to h\,\omega_0\,h^{-1} - \dot h\,h^{-1}$, 
so that 
$ \dot \J-\bigl[D,\omega_0\bigr]
\to h\,\bigl( \dot \J - \bigl[D,\omega_0\bigr]\bigr) \,h^{-1}$.

The Jacobian factor has the expected form $e^{N\times \tr({\rm something})}$, 
and thus the effective action with local counterterms 
will take the following form in the 't Hooft limit:
\begin{align}
 S=S_0[\omega_0,H] + S_{\rm loc}[\omega_0,H]
\end{align}
with%
\footnote{
There can actually be a dimension-four local counterterm, 
which corresponds to the counterterm to 
$(1/2g^2)\tr F_{\mu\nu}^2$. 
We assume that the renormalization of $g^2$ is already performed  
before the following discussions are made.
}
\begin{align}
 S_{\rm loc}
  &=\int\!\!d^3 x\,\cL_{\rm loc}(\omega_0, H,\del_\mu H)
\nn\\
 &=\int\!\!d^3 x\,N \Bigl[
  (g^2N)^0 \cL_{\rm loc}^{(3)}+(g^2N)^1 \cL_{\rm loc}^{(2)}
   +(g^2N)^2 \cL_{\rm loc}^{(1)} +(g^2N)^3 \cL_{\rm loc}^{(0)}
  \Bigr].
\end{align}
Note that the finite local counterterms $S_{\rm loc}$ 
can depend also on $\omega_0=-\dot V V^{-1}$ 
because it is invariant under time-independent (residual) gauge transformations. 
We have assigned dimensions to the fields 
as $[\omega_0]=1$, $[H]=0$ (thus $[\J]=1$), 
and demand $S_{\rm loc}$ to have the same symmetries with those of the original theory. 
That is, we require that $S_{\rm loc}$ be invariant 
under $C$, $P$ (and thus $T$) 
as well as (time-dependent) holomorphic transformations.

One can easily check that neither dimensionless nor dimension-one terms 
can contain nontrivial differential polynomials of the fields, 
so that we have $\cL_{\rm loc}^{(0)}=0$ and $\cL_{\rm loc}^{(1)}=0$.

As for dimension-two local counterterms, 
we find that 
$\cL_{\rm loc}^{(2)\,\prime}\propto \del H H^{-1}\cdot \bar\del H H^{-1}$ 
satisfies the $C$ and $P$ invariance (see Appendix \ref{JPC}). 
This term itself, however, is not invariant 
under holomorphic transformations \eq{hol_H} 
since each of the pieces $\del H H^{-1}$ and $\bar \del H H^{-1}$ 
does not transform homogeneously,
\begin{align}
 \del H H^{-1} &\to h\,\del H H^{-1}\,h^{-1} + \del h\,h^{-1},
\\
 \bar\del H H^{-1} &\to h\,\bar\del H H^{-1}\,h^{-1} 
  + H\,\bar\del \bar h\,\bar h^{-1}H^{-1}.
\end{align}
An important point is that the inhomogeneous terms, 
$\del h\,h^{-1}$ and $H\,\bar\del \bar h\,\bar h^{-1}H^{-1}$, 
belong to the zero modes of the operator 
$\ad \,\bar\del(=\bar\del)$ and $\ad D$, respectively, 
and thus, those pieces can be completed such as to transform homogeneously 
if we subtract proper zero modes from them: 
\begin{align}
 [\del H\,H^{-1}]_{\rm hol}
  \equiv \frac{1}{\bar\del}\,\bar\del\bigl(\del H\,H^{-1}\bigr),\qquad
 [\bar \del H\,H^{-1}]_{\rm hol}
  \equiv \frac{1}{\ad D}\,\ad D\bigl(\bar\del H\,H^{-1}\bigr),
\end{align}
where the inverse of an operator is defined for its part without zero modes.  
We further note that there is a subtlety for such dimension-two terms 
(related to gauge mass terms) 
because inclusion of IR regulator can also give 
a finite (generically nonlocal) term. 
Expecting that the remover of unnecessary zero modes is supplied 
by a proper IR regulator, 
we set the dimension-two ``local'' counterterms to be%
\footnote{
Another way of saying this would be 
that we introduce an IR regulator 
which removes the zero modes of the Laplacian 
$\Delta=(1/2)\bigl(\ad D\,\bar\del+\bar\del\,\ad D\bigr)$ 
from the Hilbert space. 
Note that in such Hilbert space, 
one can adopt the formal relations, 
$\bar\del^{-1}\bar\del=\bunit=\bar\del\,\bar\del^{-1}$ 
and $(\ad D)^{-1}\ad D=\bunit=\ad D\,(\ad D)^{-1}$.
}
\begin{align}
 \cL_{\rm loc}^{(2)}\equiv \frac{\beta}{8\pi^2}\,
  [\del H\,H^{-1}]_{\rm hol}\cdot [\bar \del H\,H^{-1}]_{\rm hol}
  =\frac{\beta}{8\pi^2}\,
  \frac{1}{\bar\del}\,\bar\del\bigl(\del H\,H^{-1}\bigr)\cdot
  \frac{1}{\ad D}\,\ad D\bigl(\bar\del H\,H^{-1}\bigr),
\end{align}
where $\beta$ is a certain constant to be fixed later.  
By using the definition $\J=-\del H\,H^{-1}$ and the formula 
$\delta \J=-\ad D\,(\delta H\,H^{-1})$, 
$\cL_{\rm loc}^{(2)}$ can be further rewritten as
\begin{align}
  \cL_{\rm loc}^{(2)}
  =\frac{\beta}{8\pi^2}\,
  \frac{1}{\bar\del}\,\bar\del \J\cdot \frac{1}{\ad D}\,\bar\del \J.
\end{align}

The dimension-three term is found to be
\begin{align}
 \cL_{\rm loc}^{(3)}= i\,\frac{\alpha}{2\pi}\,\,\dot H H^{-1}\cdot\bar\del \J
 \quad(\mbox{$\alpha$ is a certain constant to be fixed later})
\label{L3}.
\end{align}
This is actually invariant under $C$ and $P$ (and thus $T$) (see Appendix \ref{JPC}), 
and is also invariant under (time-dependent) holomorphic transformations 
up to total derivatives:
\begin{align}
 \dot H H^{-1}\cdot \bar\del \J
  ~\to~ \dot H H^{-1}\cdot\bar\del \J
   -2\,\bar\del\tr\bigl( \dot h\,\J\,h^{-1}\bigr)
   +2\,\del\tr\bigl(\dot{\bar h}\bar h^{-1} H^{-1}\bar\del H\bigr).  
\end{align}
Since $\dot H H^{-1}$ can be written formally as
$\dot H H^{-1}=-(\ad D)^{-1} \dot \J$, 
the local counterterm could be expressed in terms of $\J$ as
$ \cL_{\rm loc}^{(3)}
=-i\,(\alpha/2\pi)\,\,(\ad D)^{-1} \dot \J\cdot \bar\del \J$.

We thus conclude that the ``local'' counterterms are given by
\begin{align}
 \cL_{\rm loc}&=N\Bigl[\,g^2N \cL_{\rm loc}^{(2)} + \cL_{\rm loc}^{(3)} \,\Bigr]
=\beta\,\frac{g^2 N^2}{8\pi^2}\,
  \bar\del^{-1}\,\bar\del \J\cdot (\ad D)^{-1}\,\bar\del \J
  -i\alpha\,\frac{N}{2\pi}\,(\ad D)^{-1} \dot \J\cdot \bar\del \J.
\end{align}
Note that this has the expected form \eq{L_loc_guess}.
At the present stage, 
this is simply a working hypothesis 
and should be justified only by directly calculating 
the functional determinants in \eq{partition_function}, 
which we could not have done yet.

The effective Lagrangian thus becomes
\begin{align}
 \cL&=\cL_0 + \cL_{\rm loc}
\nn\\
 &= -\frac{2}{g^2}\,\omega_0\cdot\Delta\omega_0 
  +\Bigl( -\frac{2}{g^2}\,\bar\del\omega_0 
   + i\alpha\,\frac{N}{2\pi}\,(\ad D)^{-1}\bar\del \J\Bigr)
   \cdot \dot \J
\nn\\
 &~~~~+\beta\,\frac{g^2 N^2}{8\pi^2}\,
  \bar\del^{-1}\,\bar\del \J\cdot (\ad D)^{-1}\,\bar\del \J
  + \frac{2}{g^2}\,(\bar\del \J)^2.
\label{eL}
\end{align}
The canonical momenta are calculated as
\begin{align}
 \pi_V&\equiv \frac{\partial \cL}{\partial \omega_0}
 =-\frac{4}{g^2}\,\Delta\omega_0 + \frac{2}{g^2}\,\bar\del\dot \J,
\\
 \pi_\J&\equiv\frac{\partial \cL}{\partial \dot \J}
 =  -\frac{2}{g^2}\,\bar\del\omega_0 
   + i\alpha\,\frac{N}{2\pi}\,(\ad D)^{-1}\bar\del \J,
\end{align}
from which the (renormalized) effective Hamiltonian density
$\cH=\pi_V\cdot\omega_0 + \pi_\J\cdot\dot \J -\cL$ 
is obtained to be
\begin{align}
 \cH=&-\frac{g^2}{2}\,\bar\del^{-1}
  \Bigl(i\pi_\J+\alpha\,\frac{N}{2\pi}\,(\ad D)^{-1}\bar\del \J\Bigr)\cdot \cI 
\nonumber\\
  &+\alpha \,\frac{g^2N}{4\pi}\,\bar\del^{-1}\,\bar{\del }\J
   \cdot \Bigl(i\pi_\J+\alpha\,\frac{N}{2\pi}\,(\ad D)^{-1}\bar\del \J\Bigr) 
  -\beta\,\frac{g^2 N^2}{8\pi^2}\,
  \bar\del^{-1}\,\bar\del \J\cdot (\ad D)^{-1}\,\bar\del \J-\frac{2}{g^2}(\bar\del \J)^2 
\nonumber\\
   =&-\frac{g^2}{2}\,\bar\del^{-1}\,i\pi_\J\cdot \cI
   +\alpha\,\frac{g^2 N}{4\pi}\,\Bigl( (\ad D)^{-1}\bar{\del }\J\cdot {\bar{\del }}^{-1}\cI
   +{\bar{\del }}^{-1}\bar{\del }\J\cdot i\pi_\J\Bigr)
   -\frac{2}{g^2}(\bar{\del }\J)^2 \nonumber\\
   &+({\alpha}^2-\beta)\,\frac{g^2 N^2}{8\pi^2}\,\bar\del^{-1}
    \bar{\del }\J\cdot (\ad D)^{-1}\bar{\del }\J.
\label{Hamiltonian0}
\end{align}
Here $\cI^a(\bx)$ is the holomorphic Gauss-law operator that generates 
holomorphic transformations: 
\begin{align}
 \cI^a(\bx)\equiv (\ad D)^{ab}(i\pi^b_\J) (\bx)
  = (\ad D)^{ab} \Del{\J^b(\bx)}
\label{hol_Gauss_law}
\end{align}
with $(\ad D)^{ab}=\delta^{ab}\del + f^{acb}\J^c(\bx)$ . 
In Appendix \ref{HGauss} we show that
the change of variables \eq{A}--\eq{J}
leads to the following identity 
for a gauge-invariant functional 
$P[\J(\bx)]$ 
which is written with original variables as 
$P[\J(\bx)]=\hat P[A(\bx),\bar A(\bx)]$: 
\begin{align}
 \cI^a(\bx)\,P[\J]=-\bar\del\Bigl(
  \cV^{ab}(\bx)\,\frac{\delta \hat P[A,\bar A]}{\delta \bar A^b(\bx)}\Bigr),
\label{hol_Gauss_law2}
\end{align}
where $\cV^{ab}(\bx)$ is again a matrix element of $\Ad \,V(\bx)$:
$V(\bx)\,t^b\,V^{-1}(\bx)=t^a\,\cV^{ab}(\bx)$.%
\footnote{
Note that $\cV=(\cV^{ab})$ is an orthogonal matrix, 
$(\cV X)\cdot(\cV Y)=X\cdot Y$.
}
Thus, the Hamilton-Jacobi equation for the wave functional 
\begin{align}
 \Psi[\J(\bx)]=\exp\Bigl[ \frac{2\pi N}{(g^2 N)^2} 
  P[\J(\bx)] \Bigr]
\end{align}
becomes
\begin{align}
 \frac{g^2}{2}E=
  \int\Bigl[ &-\Bigl( \frac{\alpha}{4}\,(\ad D)^{-1}\bar\del \J
   + \frac{1}{4m^2}\,\frac{\delta P}{\delta \J}\Bigr)\cdot
     \cV\,\frac{\delta \hat P}{\delta \bar A}
   + \frac{\alpha}{4}\,{\bar{\del }}^{-1}\bar{\del }\J\cdot \frac{\delta P}{\delta \J}
   -\bigl(\bar\del \J\bigr)^2 \nn\\
 &+({\alpha}^2-\beta)\,\frac{m^2}{4}\,{\bar{\del }}^{-1}
   \bar{\del }\J\cdot (\ad D)^{-1}\bar{\del }\J
 \Bigr] 
 \qquad(m=g^2 N/2\pi).
\label{fh}
\end{align}
In Appendix \ref{HGauss} we further show that 
the operators 
\begin{align}
 &\bar \cN\equiv -\int(\ad D)^{-1}\bar\del \J\cdot \cV\Del{\bar A}
  =-\int_{\bx,\by}\bigl((\ad D)^{-1}\bigr)^{ab}(\bx,\by)\,\bar\del \J^b(\by)
    \, \cV^{ac}\Del{\bar A^c(\bx)}
\label{N-bar}
\\
 &\cN \equiv \int{\bar{\del }}^{-1}\bar{\del }\J\cdot \frac{\delta}{\delta \J}
\end{align}
count the number of $\bar\del$ and $D$, respectively, in the gauge-invariant functional 
on which they act. 
That is, when $P[\J]$ is expanded as a formal Laurent series in $\bar\del$ and $D$, 
the operators $\bar\cN$ and $\cN$ act linearly on each term in the expansion 
and return its orders in $\bar\del$ and $D$, respectively. 
For example, on $P[\J]=\sum_n\,c_n\,\int \bar\del \J\cdot (\ad D)^n\bar\del \J$, 
they act as 
$\bar\cN P[\J]=\sum_n 2 c_n\int \bar\del \J\cdot (\ad D)^n\bar\del \J$ 
and $\cN P[\J]=\sum_n\,(n+2) c_n\int \bar\del \J\cdot (\ad D)^n\bar\del \J$. 
We have used the relation $\bar\del \J=[\bar\del, D]$. 
Note that the operators $\bar\cN$ and $\cN$ are parity-conjugates to each other 
and have a definite meaning that is invariant under holomorphic transformations.

One can see that the last term in \eq{fh} leads to an IR divergence
in the gauge invariant sector 
since $\bar \del \J$ represents the field-strength, 
$\bar\del \J=(-i/2)\,\V F_{12}\V^{-1}$.
We thus set $\beta=\alpha^2$ 
and drop the term in the following discussions. 
The final form of the Hamiltonian density and the corresponding H-J equation thus become%
\footnote{
Another possibility for dimension-three local counterterm other than \eq{L3} could be 
to take the WZ-like term : 
\begin{align}
 \cL_{\rm loc}^{(3)} = {\rm const}\,\tr \bigl(\dot H\,H^{-1}\,
  \bigl[ \del H\,H^{-1}, \bar\del H\,H^{-1}\bigr] \bigr).
\nn
\end{align}
This is actually invariant under (time-dependent) holomorphic transformations 
because infinitesimal holomorphic transformations 
(actually any local changes of variable) give total derivatives. 
One can easily check that this is also invariant under $C$ and $P$ 
(and thus $T$) (see Appendix \ref{JPC}). 
We expect that it yields the same Hamiltonian with \eq{ehf1} 
when rewritten in terms of $\J$
together with a proper IR cutoff corresponding to this local counterterm.
}
\begin{align}
\cH=&-\frac{g^2}{2}\bar{\del}^{-1}\,i\pi_\J\cdot \cI
   +\alpha\,\frac{g^2 N}{4\pi}\,
     \Bigl( (\ad D)^{-1}\bar{\del }\J\cdot {\bar{\del }}^{-1}\cI
   +{\bar{\del }}^{-1}\bar{\del }\J\cdot i\pi_\J\Bigr)
   -\frac{2}{g^2}\,(\bar{\del }\J)^2,  \label{ehf1}\\
 \frac{g^2}{2}E=&
  \frac{\alpha}{4}\, \bar \cN P
  +\frac{\alpha}{4}\, \cN P
   - \int\Bigl[ \frac{1}{4m^2}\,\frac{\delta P}{\delta \J}\cdot
     \cV\,\frac{\delta \hat P}{\delta \bar A}
   +\bigl(\bar\del \J\bigr)^2 \Bigr]
 \qquad(m=g^2 N/2\pi).
\label{S_eqn}
\end{align}

We conclude this section with a comment on the value of $\alpha$. 
The preceding discussions do not fix the actual value of $\alpha$. 
For example, the removal of the IR divergence 
can be performed for any value of $\alpha$ (when $\beta=\alpha^2$). 
Although $\alpha$ should be determined by directly evaluating 
the functional determinants in \eq{partition_function}, 
this can be fixed by comparing our local counterterms 
with the finite corrections obtained by Karabali and Nair 
in their original Hamiltonian formalism \cite{KKN}. 
To see this, we first notice that in spin-$0$ wave functionals,
$\bar\del$ always appears 
with the combination $\bar\del \J=[\bar\del, D]$ or in the form of 
$\Delta=(1/2)\bigl(\ad D\,\bar\del + \bar\del\,\ad D\bigr)$. 
Thus, our operator $\bar\cN$ can also be interpreted 
as counting the number of $D$ (i.e.\ $\bar \cN=\cN$), 
or the number of $\J$ if we neglect $\del$ in $D=\del+\J$ 
in the IR limit;  
$\bar\cN = \cN \sim \int \J\cdot \delta/\delta \J~\,\mbox{(IR)}$.
Then \eq{S_eqn} in the IR regime agrees with 
the one originally obtained by Karabali and Nair 
using the heat kernel regularization \cite{KKN} 
if we take $\alpha=1$. 
We set $\alpha=\beta=1$ in the following argument.

\section{Ground state wave functional}
\label{ground_state}

In this section, we solve the H-J equation for the ground state 
wave functional 
$\Psi[\J(\bx)]=\exp\Bigl[\bigl( 2\pi N/(g^2N)^2\bigr)\,P[\J(\bx)]\bigr]$,
\begin{align}
 0= \frac{1}{4}\, \bar \cN P
    +\frac{1}{4}\, \cN P
   - \int\Bigl[ \frac{1}{4m^2}\,\frac{\delta P}{\delta \J}\cdot
     \cV\,\frac{\delta \hat P}{\delta \bar A}
   +\bigl(\bar\del \J\bigr)^2 \Bigr].
\label{S_eqn2}
\end{align}

\subsection{Preparations}

We expand the functional $P[\J]$ as
\begin{align}
 P[\J]=\sum_{n=2}^\infty \frac{1}{m^{2(n-2)}}P_n[\J]. 
\end{align}
Here $P_n[\J]\ (n\geq 2)$ is a holomorphic invariant functional 
whose order in $\J$ is at least $n$.
For example, 
\begin{align}
 P_2[\J]&=\int 
  \bar\del \J\cdot K\Bigl(\frac{\Delta}{m^2}\Bigr) \bar\del \J 
  =\int \Bigl[\,\bar\del \J\cdot K\Bigl(\frac{\del\bar\del}{m^2}\Bigr)\,
   \bar\del \J
   + O(\J^3)\Bigr]
\end{align}
is the Gaussian part
augmented with higher order terms in $\J$ 
such that it becomes holomorphic invariant.
This can be rewritten in terms of the original gauge potentials as
\begin{align}
 P_2[\J]=\int 
  [ D_{\bar z},D_z]\cdot K\Bigl(\frac{\hat\Delta}{m^2}\Bigr)
  [ D_{\bar z},D_z]
  \,\equiv\, \hat P_2[A,\bar A],
\end{align}
where $\hat\Delta$ is the Laplacian with respect to 
the original gauge potentials: 
\begin{align}
 \hat\Delta \equiv \frac{1}{2}\bigl(
  \ad D_z\,\ad D_{\bar z} + \ad D_{\bar z}\,\ad D_z\bigr)
  = (\Ad \V)^{-1}\,\Delta\,\,\Ad \V.
\end{align}

Since $P_n[\J]$ is holomorphic invariant, 
it must be constructed only from $\bar\del \J$ 
and $\Delta =(1/2)\bigl(\bar\del D+D\bar\del \bigr)$, 
and thus, when $\delta /\delta \J$ acts on $P_n[\J]$, 
we only need to consider two cases:
one is when it hits on $\bar\del \J$ in $P_n[\J]$ 
and the other is on $\Delta$. 
Both cases can be easily dealt with 
by using the following formulas: 
\begin{align}
 \frac{\delta }{\delta \J}\int \bar\del \J\cdot X
  & =-\bar\del X \quad\Bigl(
    \equiv -\bar\del \frac{\delta }{\delta \bar\del \J}\int \bar\del \J\cdot X
   \Bigr),
\label{variation01}\\
 \frac{\delta }{\delta \J}\int \Delta X\cdot Y
  &=\frac{1}{2}\bigl(\, \bigl[\bar\del X,Y\bigr]-\bigl[X,\bar\del Y\bigr]\, \bigr),
\label{variation03}
\end{align}
the latter of which can be derived as
\begin{align}
 \delta_\J\int \Delta X\cdot Y
  =\frac{1}{2}\int (\,\ad\delta \J\cdot \bar\del+\bar\del \cdot \ad\delta \J\,)X\cdot Y
  =\int\delta \J\cdot \frac{1}{2}\bigl(\bigl[\bar\del X,Y\bigr]-\bigl[X,\bar\del Y\bigr] \bigr).
\end{align}
Note that the former case (when $\delta/\delta \J$ hits on $\bar\del \J$) 
necessarily reduces the order of the functional in $\J$ by one, 
while the latter (when $\delta/\delta \J$ hits on $\Delta$) does not 
change the order. 
This observation leads to the formula:
\begin{align}
 \frac{\delta P_n}{\delta \J}
  &=\Bigl[\frac{\delta P_n}{\delta \J}\Bigr]_{n-1}
    +\Bigl[\frac{\delta P_n}{\delta \J}\Bigr]_n
  =-\bar\del\frac{\delta P_n}{\delta \bar\del \J}
  +\Bigl[\frac{\delta P_n}{\delta \J}\Bigr]_{n}.
\label{P-J}
\end{align}
Similar analysis can be made for $\cV\,\delta/\delta\bar A$ 
acting on $P[\J]=\hat P[A,\bar A]$. 
We then obtain the following formulas: 
\begin{align}
 \cV \frac{\delta }{\delta \bar A}\int \bar \del \J\cdot X
  &=\cV \frac{\delta }{\delta \bar A}\int \bigl[D_{\bar z},D_z\bigr]\cdot \hat X=\ad D\,X
  \quad\Bigl(\equiv \ad D\frac{\delta }{\delta \bar\del \J}
   \int \bar\del \J\cdot X\Bigr),
\label{variation02}\\
 \cV \frac{\delta }{\delta \bar A}\int \Delta X\cdot Y
  &=\cV \frac{\delta }{\delta \bar A}\int \hat \Delta \hat X\cdot \hat Y
  =\frac{1}{2}\bigl( \bigl[\ad D\, X,Y\bigr]-\bigl[X,\ad D\, Y\bigr] \bigr)\label{variation04},
\end{align}
which are derived as
\begin{align}
 \delta _{\bar A}\int \bigl[D_{\bar z},D_z\bigr]\cdot \hat X
  &=\int \bigl[\delta \bar A, D_z\bigr]\cdot\hat X
  =\int \delta  \bar A \cdot\bigl[D_z, \cdot\hat X\bigr],
\\
 \delta_{\bar A}\int \hat \Delta \hat X\cdot \hat Y
  &=\frac{1}{2}\int 
  (\,\ad\delta \bar A \,\ad D_z+\,\ad D_z \,\ad\delta \bar A\,)\hat X\cdot \hat Y
\nn\\
 &=\int\delta \bar A \cdot \frac{1}{2}
  \Bigl(\bigl[\ad D_{z}\hat X,\hat Y\bigr]-\bigl[\hat X, \ad D_{z}\hat Y\bigr]\Bigr).
\end{align}
We also have a formula similar to \eq{P-J},
\begin{align}
 \cV \frac{\delta P_n}{\delta \bar A}
  =\Bigl[\cV \frac{\delta P_n}{\delta \bar A}\Bigr]_{n-1}
   +\Bigl[\cV \frac{\delta P_n}{\delta \bar A}\Bigr]_n
  =\ad D\frac{\delta P_n}{\delta \bar\del \J}
  +\Bigl[\cV \frac{\delta P_n}{\delta \bar A}\Bigr]_{n}.
\end{align}
For example, on $P_2[\J]=\sum (K_n/m^{2n})\cO_n$ 
with 
$\cO_n\equiv \int \bar\del \J\cdot {\Delta}^n\bar\del \J$, 
the operators $\delta/\delta \J$ and $\cV\,\delta/\delta\bar A$ 
act as (setting $L=\Delta/m^2$):
\begin{align}
 &\frac{\delta P_2}{\delta \J}
  =\Bigl[\frac{\delta P_2}{\delta \J}\Bigr]_{1}
  +\Bigl[\frac{\delta P_2}{\delta \J}\Bigr]_{2}
\nn\\
 &\mbox{with}\quad\Bigl[\frac{\delta P_2}{\delta \J}\Bigr]_{1}
  =-2\,\bar\del K(L)\,\bar\del \J,
  \quad
  \Bigl[\frac{\delta P_2}{\delta \J}\Bigr]_{2}
  =\sum_{n=0}^\infty\sum_{m=0}^{n-1}\frac{K_n}{m^{2n}}\,
  \bigl[\bar\del \Delta^m\bar\del \J,\Delta^{n-m-1}\bar\del \J\bigr],
\label{variP2-1}
\end{align}
and
\begin{align}
 &\cV\frac{\delta \hat P_2}{\delta \bar A}
  =\Bigl[\cV\frac{\delta \hat P_2}{\delta \bar A}\Bigr]_{1}
  +\Bigl[\cV\frac{\delta \hat P_2}{\delta \bar A}\Bigr]_{2}
\nn\\
 &\mbox{with}\quad
  \Bigl[\cV\frac{\delta \hat P_2}{\delta \bar A}\Bigr]_{1}
  =2 \,\ad D\, K(L)\,\bar\del \J,
  \quad
  \Bigl[\cV\frac{\delta \hat P_2}{\delta \bar A}\Bigr]_{2}
  = \sum_{n=0}^\infty\sum_{m=0}^{n-1}\frac{K_n}{m^{2n}}\,
   \bigl[\ad D\Delta^m\bar\del \J,\Delta^{n-m-1}\bar\del \J\bigr].
\label{variP2-2}
\end{align}

\subsection{Gaussian part}

After making rather lengthy preparations presented above, 
we are now in a position to calculate 
$P_n[\J]$ $(n\geq 2)$.
First, by using \eq{variP2-1} and \eq{variP2-2}, 
the holomorphic-invariant $O(\J^2)$ part in \eq{S_eqn2} is expressed as
\begin{align}
 0&=\frac{1}{4}\, \bar \cN P_2
    +\frac{1}{4}\, \cN P_2
   - \frac{1}{4m^2}\int\Bigl[ \frac{\delta P_2}{\delta \J}\Bigr]_{1}\cdot
   \Bigl[ \cV\frac{\delta \hat P_2}{\delta \bar A}\Bigr]_{1}
   -\int(\bar\del \J)^2 + O(\J^3)
\nonumber\\
  &=\frac{1}{4}\, \bar \cN P_2
    +\frac{1}{4}\, \cN P_2
   - \int\bar\del \J\cdot L\,K^2(L)\,\bar\del \J
   -\int(\bar\del \J)^2 + O(\J^3).
\label{HJ-2}
\end{align}
For $K(L)=\sum_n K_n L^n$ with $L=\Delta/m^2$, 
$\bar\cN$ acts as%
\footnote{
This gives a justification for the rule given by 
Leigh et al.\ (see Appendix A of \cite{LMY}), 
where \eq{N-bar_K} is assumed to hold 
as a consequence of complicated normal orderings 
in subject to successive variations in the Schr\"odinger equation. 
Note that their operator $\int \!\J\cdot \delta/\delta \J$ 
is replaced by $(\bar \cN+\cN)/2$ in our framework, 
where we assume that any necessary normal orderings 
are already taken care of by adding local counterterms to Lagrangian. 
}
\begin{align}
 \bar\cN \,\int \bar\del \J\cdot K(L)\,\bar\del \J
  &=\bar\cN\,\sum_n K_n \int \bar\del \J \cdot
   \Bigl(\frac{\Delta}{m^2}\Bigr)^{\! n}\,\bar\del \J
  =\sum_n (n+2) \,K_n \int \bar\del \J\cdot
   \Bigl(\frac{\Delta}{m^2}\Bigr)^{\! n}\,\bar\del \J 
\nn\\
 &= \int \bar\del \J\cdot \Bigl(L \frac{d}{dL}+2\Bigr)K(L)\,\bar\del \J.
\label{N-bar_K1}
\end{align}
The action of $\cN $ is the same with that of $\bar{ \cN}$
for functionals of spin zero. 
\begin{align}
 \cN \,\int \bar\del \J\cdot K(L)\,\bar\del \J
  = \int \bar\del \J\cdot \Bigl(L \frac{d}{dL}+2\Bigr)K(L)\,\bar\del \J.
\label{N-bar_K}
\end{align}
Thus, collecting the coefficients of $\J^2$ terms in \eq{HJ-2}, 
we obtain the differential equation for the kernel $K(L)$,
\begin{align}
 &\Bigl(\frac{1}{2}\,L\frac{d}{dL}+1\Bigr)K(L) - L K^2(L) -1 = 0,
\label{LMYeq}
\end{align}
which certainly reproduces the result of Leigh et al.\ \cite{LMY}.

Once the equation \eq{LMYeq} is obtained, 
one can derive all the results in \cite{LMY} with the same logic. 
First, the equation has a normalizable solution   
\begin{align}
  K(L) = \frac{1}{\sqrt{L}}\,\frac{J_2(4\sqrt{L})}{J_1(4\sqrt{L})},
\label{kl}
\end{align}
which has the following asymptotic form:
\begin{align}
 K(L) 
   = \left\{
      \begin{array}{cl}
       1 + (2/3)L + \cdots & (L\to 0\mbox{: IR})\\
       1/\sqrt{-L} & (L\to -\infty\mbox{: UV})\,.
      \end{array}
      \right.
\end{align}
Thus, the ground-state wave functional behaves in the IR and UV regions as
\begin{align}
 \Psi_{\rm IR}
  &=\exp\Bigl[ \,\frac{2\pi N}{(g^2 N)^2} \int_\bx 
    (\bar\del \J(\bx))^2 +\cdots\Bigr]
\nn\\
 &=\exp\Bigl[\,-\,\frac{\pi}{2 g^4N}
   \int_{\bx} (F_{12}(\bx))^2+\cdots\Bigr],
\\
 \Psi_{\rm UV}
  &=\exp\Bigl[ \,\frac{2}{g^2} \int_\bx 
    \bar\del \J\cdot\frac{1}{\sqrt{-\del_\bx^2}}\,\bar\del \J 
    +\cdots\Bigr]
\nn\\
 &=\exp\Bigl[\,-\,\frac{1}{2 g^2}
   \int_{\bx} F_{12}(\bx)\cdot \frac{1}{\sqrt{-\del_\bx^2}}\,F_{12}(\bx)
    +\cdots\Bigr].
\end{align}
The UV wave functional reproduces that of massless free gluons. 
The vev of a large Wilson loop, on the other hand, 
can be evaluated with the use of $\Psi_{\rm IR}$ as 
\begin{align}
 \Bigl< \tr {\rm P} e^{-\oint A} \Bigr>
  &\sim \int\![dA(\bx)d\bar A(\bx)] \,\tr {\rm P} e^{-\oint\! A} 
  \cdot \bigl| \Psi_0^{\rm (IR)}\bigr|^2 \nn\\
 &= \int\![dA(\bx) d\bar A(\bx)] \,\tr {\rm P} e^{-\oint\! A} 
   \cdot e^{-\,(\pi/g^4N)
   \int_{\bx}\!(F_{12}(\bx))^2+\,\cdots}.
\end{align}
This is nothing but the Wilson loop average of 2D YM 
with 2D gauge coupling $g_2^2 = \,g^4 N/2\pi = m \,g^2$. 
As was first pointed out by Karabali et al.\ \cite{KKN}, 
the string tension of 3D pure YM theory is then calculated 
from the known value of the string tension of 2D YM theory \cite{2D} 
as
$
 \sigma=g_2^2\,(N^2-1)/4N= g^4\,(N^2-1)/8\pi
$, i.e.
\begin{align}
 \sqrt{\sigma}=\frac{1}{\sqrt{8\pi}}\,g^2 N 
  =\sqrt{\frac{\pi}{2}}\,m
  \quad(N\to\infty).
\end{align}
The computation of 3D glueball masses could be carried out 
in the same way as in \cite{LMY}.

\subsection{Higher-order corrections}

As for the $O(\J^3)$ part, we have the equation
\begin{align}
 0&= \frac{1}{2}\bar \cN P_3+\frac{1}{2}\cN P_3
\nn\\
 &~~~-\frac{1}{2}\int\Bigl( \Bigl[ \frac{\delta P_2}{\delta \J}\Bigr]_{1}
    \cdot\Bigl[ \cV\frac{\delta \hat P_2}{\delta \bar A}
   +\frac{1}{m^2}\cV\frac{\delta \hat P_3}{\delta \bar A}\Bigr]_{2}
   +\Bigl[ \cV\frac{\delta \hat P_2}{\delta \bar A}\Bigr]_{1}
    \cdot \Bigl[ \frac{\delta P_2}{\delta \J}
   +\frac{1}{m^2}\frac{\delta P_3}{\delta \J}\Bigr]_{2}
   \Bigr) 
  + O(\J^4)
\nonumber\\
  &= \frac{1}{2}\bar \cN P_3+\frac{1}{2}\cN P_3
\nn\\
 &~~~
   -\int\bigl( -\bar\del K(L)\,\bar\del \J\bigr)\cdot 
   \Bigl( \sum_{n=0}^\infty\sum_{m=0}^{n-1}\frac{K_n}{m^{2n}}
   \bigl[\ad D\,\Delta^m\bar\del \J,\Delta^{n-m-1}\bar\del \J\bigr]
   +\frac{1}{m^2}\ad D\frac{\delta P_3}{\delta \bar\del \J}
   \Bigr) 
\nonumber\\
   &~~~-\int\bigl( \ad D\, K(L)\,\bar\del \J\bigr)
   \cdot \Bigl( \sum_{n=0}^\infty\sum_{m=0}^{n-1}\frac{K_n}{m^{2n}}
   \bigl[\bar\del \Delta^m\bar\del \J,\Delta^{n-m-1}\bar\del \J\bigr]
   -\frac{1}{m^2}\bar\del \frac{\delta P_3}{\delta \bar\del \J}
   \Bigr) + O(\J^4)
\nonumber\\
   &= \frac{1}{2}\bar \cN P_3+\frac{1}{2}\cN P_3
   -\int \Bigl(2K(L)\bar\del \J
   \cdot L\frac{\delta P_3}{\delta \bar\del \J}
   +\sum_{k=0}^\infty\sum_{l=0}^\infty\sum_{m=0}^{l-1}
    \frac{K_kK_l}{m^{2(k+l)}}\cO_{k,m,l-m-1}
   \Bigr)  + O(\J^4),
\label{HJ-3}
\end{align}
where we have introduced the operators
\begin{align}
 \cO_{k,l,m}&\equiv
  \int\Delta^k\bar\del \J
  \cdot\bigl( \bar\del \bigl[\ad D\Delta^l\bar\del \J,\Delta^m\bar\del \J\bigr]
  +\ad D\bigl[\Delta^m\bar\del \J,\bar\del \Delta^l\bar\del \J\bigr]\bigr)
  \quad(k,l,m\geq 0)\label{oklm}.
\end{align}
They are totally symmetric with respect to their indices up to $O(\J^4)$. 
Actually,
\begin{align}
\cO_{k,l,m}&=\int\Delta^k\bar\del \J
  \cdot\bigl(\bar\del \bigl[\ad D\Delta^l\bar\del \J,\Delta^m\bar\del \J\bigr]
  +\ad D \bigl[\Delta^m\bar\del \J,\bar\del \Delta^l\bar\del \J\bigr]\bigr)
\nonumber\\
 &=\int\Delta^k\bar\del \J\cdot
  \bigl(\bigl[\ad D\Delta^l\bar\del \J,\bar\del \Delta^m\bar\del \J\bigr]
  +\bigl[\ad D\Delta^m\bar\del \J,\bar\del \Delta^l\bar\del \J\bigr]\bigr)   +O(\J^4)
\nonumber\\
 &=\int\Delta^k\bar\del \J\cdot 
  \bigl(\bigl[\ad D\Delta^m\bar\del \J,\bar\del \Delta^l\bar\del \J\bigr]
  +\bigl[\ad D\Delta^l\bar\del \J,\bar\del \Delta^m\bar\del \J\bigr]\bigr)  +O(\J^4)
\nonumber\\
 &=\int\Delta^k\bar\del \J
  \cdot \bigl(\bar\del \bigl[\ad D\Delta^m\bar\del \J,\Delta^l\bar\del \J\bigr]
  +\ad D\bigl[\Delta^l\bar\del \J,\bar\del \Delta^m\bar\del \J\bigr]\bigr)  +O(\J^4)
\nonumber\\
 &=\cO_{k,m,l}+O(\J^4),
\\
 \cO_{k,l,m}&=\int\Delta^k\bar\del \J
  \cdot\bigl(\bar\del \bigl[D\Delta^l\bar\del \J,\Delta^m\bar\del \J\bigr]
  +\ad D\bigl[\Delta^m\bar\del \J,\bar\del \Delta^l\bar\del \J\bigr]\bigr)
\nonumber\\
 &=\int \Bigl( \ad D\Delta^l\bar\del \J
  \cdot\bigl[\bar\del \Delta^k\bar\del \J,\Delta^m\bar\del \J\bigr]
  +\bar\del \Delta^l\bar\del \J\cdot \bigl[-\ad D\Delta^k\bar\del \J,
  \Delta^m\bar\del \J\bigr]\Bigr)
\nonumber\\
 &=\int \Delta^l\bar\del \J\cdot \bigl( \ad D\bigl[\Delta^m\bar\del \J,
  \bar\del \Delta^k\bar\del \J\bigr]
  +\bar\del \bigl[\ad D\Delta^k\bar\del \J,\Delta^m\bar\del \J\bigr]\bigr)
\nonumber\\
 &=\cO_{l,k,m}.         
\end{align}

Equations \eq{HJ-3} and \eq{oklm} motivate us 
to set $P_3$ to be in the following form :
\begin{align}
 P_3=\sum_{klm}\frac{d_{k,l,m}}{m^{2(k+l+m+1)}}\cO_{k,l,m}
\label{P3a}
\end{align}
with totally symmetric coefficients $d_{k,l,m}$. 
Then, up to $O(\J^4)$ terms, 
each part in \eq{HJ-3} is rewritten into the following form:
\begin{align}
 \cN P_3=\bar \cN P_3&=\sum_{klm}\frac{(k+l+m+4)}{m^{2(k+l+m+1)}}d_{k,l,m}\cO_{k,l,m}, 
\end{align}
\begin{align}
 \int K(L)\,\bar\del \J\cdot L\frac{\delta P_3}{\delta \bar\del \J} 
  &=\sum_n\sum_{k,l,m}\frac{K_n d_{k,l,m}}{m^{2(n+k+l+m+2)}}
  \int {\Delta }^n\bar\del \J \cdot
  \Delta \frac{\delta \cO_{k,l,m}}{\delta \bar\del \J} 
\nonumber\\
 &=\sum_n\sum_{k,l,m}\frac{K_n d_{k,l,m}}{m^{2(n+k+l+m+2)}}
  \bigl(  \cO_{k+n+1,l,m}+\cO_{k,l+n+1,m}+\cO_{k,l,m+n+1} \bigr)
\nonumber\\
 &=\sum_n\sum_{k,l,m}\frac {K_n}{m^{2(k+l+m+1)}}
  \bigl(d_{k-1-n,l,m}+d_{k,l-1-n,m}+d_{k,l,m-1-n}\bigr)\cO_{k,l,m},
\end{align}
\begin{align}
 \sum_r\sum_n\sum_{m=0}^{n-1}\frac{K_rK_n}{m^{2(r+n)}}\cO_{r,n-1-m,m}
  &=\sum_{klm}\frac{K_k K_{l+m+1}}{m^{2(k+l+m+1)}}\cO_{k,l,m}
\nonumber\\
 &=\frac{1}{3}\sum_{k,l,m}
  \frac{(K_k K_{l+m+1}+K_l K_{m+k+1}+K_m  K_{k+l+1})}{m^{2(k+l+m+1)}}\cO_{k,l,m}.
\end{align}
Collecting the coefficients of $\cO_{k,l,m}$, 
we obtain the recursion equations for $d_{k,l,m}$ :
\begin{align}
 0&=(k+l+m+4)d_{k,l,m}
  -2\Bigl(
   \sum_{n} K_n d_{k-1-n,\,l,\,m}
   +\sum_{n} K_n d_{k,\,l-1-n,\,m}
   +\sum_{n} K_n d_{k,\,l,\,m-1-n} \Bigr)
\nonumber\\ 
 &~~-\frac{1}{3}\bigl(K_k K_{l+m+1}+K_l K_{m+k+1}+K_m K_{k+l+1}\bigr),
\end{align}
which can actually be solved recursively 
for the increasing ordering in  the sum of indices, $k+l+m$, 
with the initial condition $d_{0,0,0}=(1/4)K_1=1/6$. 
One can easily check that 
this recursion equations are equivalent to the following differential equation 
for the generating function 
$D(L_1,L_2,L_3)\equiv \sum_{k,l,m}d_{k,l,m}{L_1}^k{L_2}^l{L_3}^m$:
\begin{align}
&\Bigl[ L_1\Bigl( \frac{\del }{\del L_1}-2K(L_1)\Bigr)
+L_2\Bigl( \frac{\del }{\del L_2}-2K(L_2)\Bigr)
+L_3\Bigl( \frac{\del }{\del L_3}-2K(L_3)\Bigr)
+4\Bigr]D(L_1,L_2,L_3)\nonumber\\
&=\frac{1}{3}\Bigl( K(L_1)\frac{K(L_2)-K(L_3)}{L_2-L_3}
+K(L_2)\frac{K(L_3)-K(L_1)}{L_3-L_1}
+K(L_3)\frac{K(L_1)-K(L_2)}{L_1-L_2}\Bigr).
\end{align}
Calculation of $O(\J^4)$ part can also be performed 
in a similar way.

We conclude this section with a comment on the glueball mass 
calculations carried out by Leigh et al.\ \cite{LMY},
who set the Ansatz that the vacuum wave functional is Gaussian. 
Although the obtained results are in excellent agreement 
with lattice data \cite{LTW}, 
we have seen above 
that there exist nonvanishing corrections to the Gaussian part, 
so that there must be some explanation why the Gaussian Ansatz 
gives a good approximation. 
We still do not have a satisfying answer for that yet, 
but we expect that higher-order corrections will take the form 
of integrating a rational function of Bessel functions, 
which will get suppressed strongly due to the cancellation in oscillating integrals.

\section{Conclusions and outlook}

In this paper we proposed a new method 
to analyze 3D Yang-Mills theory in the 't Hooft limit 
based on the effective Lagrangian and holomorphic symmetry. 
We wrote down the equation which determines the wave functional, 
and showed that it can be solved recursively in $\J$. 
We gave sample calculations for the Gaussian and $O(\J^3)$ parts 
of the ground state wave functional, 
and showed that the Gaussian part  
reproduces the result obtained by Leigh et al.\ \cite{LMY}.

There are many things that should be carried out. 
First of all, we need to elaborate the analysis itself, 
especially to further clarify the origin 
of the local counterterms 
and to determine the coefficient $\alpha$ in \eq{S_eqn} 
by directly evaluating the functional determinants 
in \eq{partition_function}.

Other possible directions are to consider finite size effects 
in the present framework, 
and also to include many other fields including 
scalars and fermions in various representations. 
Among them, the inclusion of adjoint scalars 
should be particularly interesting, 
since it can be used to construct 4D YM theory via deconstruction \cite{ACG}. 
Investigations along the above line is now in progress 
and will be reported in our future communication \cite{FK2}.


\section*{Acknowledgments}

The authors thank Antal Jevicki, Tetsuya Onogi, Tadashi Takayanagi 
and Alyosha Zamolodchikov 
for useful discussions. 
A preliminary result of part of the present article 
was announced by M.F.\ at the YITP workshop 
YITP-W-07-05 on ``String Theory and Quantum Field Theory,'' 
August 6-10, 2007.
He thanks the participants, 
especially Hiroshi Suzuki, for useful discussions. 
This work was supported in part by the Grant-in-Aid for 
the 21st Century COE
``Center for Diversity and Universality in Physics'' 
from the Ministry of Education, Culture, Sports, Science 
and Technology (MEXT) of Japan.  
M.F.\ is also supported by the Grant-in-Aid for 
Scientific Research No.\ 19540288 from MEXT.

\appendix

\section{$J^{PC}$}
\label{JPC}

We summarize the quantum numbers assigned to fields.

~

\noindent
\underline{$\bJ$}:

A field $\Phi(z,\bar z)$ is said to have spin $J$ when it transforms as
\begin{align}
   \Phi' (z, \bar z) = e^{iJ\theta}\,\Phi (e^{i\theta}z,e^{-i\theta}\bar z).
\end{align}
This definition leads to the following assignment:
\begin{center}
\begin{tabular}{c|cccccc|ccc}
   & $A$  & $\bar A$ & $\del$ & $\bar\del$ & $H$ & $\J$ 
   & $\bar\del \J$ & $D$  & $\Delta$ \\ \hline
     $J$ (spin) & $+1$ & $-1$     & $+1$   & $-1$       & $0$ & $+1$ & $0$
         & $+1$ & $0$
\end{tabular}  
\end{center}

\noindent
\underline{$\bP$}:

Parity transformation is defined as $(x^1,x^2)~\stackrel{P}\to~(x^1,-x^2)$ 
or
\begin{align}
 x=(z,\bar z,t) ~\stackrel{P}\to~ 
  x_P\equiv(\bar z,z,t), 
\end{align}
under which the original gauge potentials $A(x)$ and $\bar A(x)$ 
transform as 
\begin{align}
 A(x) &~\stackrel{P}{\to}~ \bar A(x_P) 
\\
 \bar A(x) &~\stackrel{P}{\to}~ A(x_P) 
.
\end{align}
This gives 
\begin{align}
 V(x) ~\stackrel{P}\to~ V^\dag{}^{-1}(x_P)
  \quad \Rightarrow \quad 
  H(x) ~\stackrel{P}\to~ H^{-1}(x_P)
\end{align}
and
\begin{align}
 \omega_0(x) &~\stackrel{P}\to~ 
  +\,\bigl(H^{-1}\,(\omega_0+\dot H\,H^{-1})\,H\bigr)(x_P)
\\
 \J(x) &~\stackrel{P}\to~ +\,\bigl(H^{-1}\,\bar\del H\bigr)(x_P)
\\
 \bar\del \J(x) &~\stackrel{P}\to~ 
    -\,\bigl(H^{-1}\, \bar\del \J\, H \bigr)(x_P)
,
\end{align}
which show that the matrix $\dot H H^{-1}\bar\del \J$ transforms as
\begin{align}
 \dot H H^{-1} \bar\del \J(x) ~\stackrel{P}\to~ 
  +\,\bigl(H^{-1} \dot H H^{-1} \bar\del \J\,H\bigr)(x_P).
\end{align}

\noindent
\underline{$\bC$}:

Charge conjugation is defined for the original gauge potentials 
$A(x)$ and $\bar A(x)$ as 
\begin{align}
 A_\mu(x) &~\stackrel{C}\to~ -\,A_\mu^{\rm T}(x) 
\\
 \bar A_\mu(x) &~\stackrel{C}\to~ -\, \bar A_\mu^{\rm T}(x) 
.
\end{align}
This gives
\begin{align}
 V(x) ~\stackrel{C}\to~ (V^{\rm T}){}^{-1}(x)
  \quad \Rightarrow \quad 
  H(x) ~\stackrel{C}\to~ (H^{\rm T}){}^{-1}(x)
\end{align}
and
\begin{align}
 \omega_0(x) &~\stackrel{C}\to~ -\,\omega_0^{\rm T}(x)
\\
 \J(x) &~\stackrel{C}\to~ -\,\J^{\rm T}(x)
\\
 \bar\del \J(x) &~\stackrel{C}\to~ 
    -\,\bar\del \J^{\rm T}(x)
,
\end{align}
which show that the matrix $\dot H H^{-1}\bar\del \J$ transforms as
\begin{align}
 \dot H H^{-1} \bar\del \J(x) ~\stackrel{C}\to~ 
 \bigl(\bar\del \J\,\dot H H^{-1}\bigr){}^{\rm T}(x).
\end{align}

\section{Holomorphic Gauss-law operator $\cI^a(\bx)$}
\label{HGauss}

We collect properties of the holomorphic Gauss-law operator 
defined in \eq{hol_Gauss_law}, 
\begin{align}
 \cI^a(\bx)\equiv 
  (\ad D)^{ab} \Del{\J^b(\bx)},
\end{align}
where $(\ad D)^{ab}=\delta^{ab}\del + f^{acb}\J^c(\bx)$. 
The following statements actually hold for arbitrary finite $N$.

We first show that under the change of variables \eq{A}--\eq{J}
the variations transform as%
\footnote{
In text, $\ad\,\bar\del$ is usually abbreviated as $\bar\del$. 
In the following, the variation with respect to $H$ is always rewritten 
in terms of $\J$ by using the relation $\delta \J=-\ad D\,(\delta H\,H^{-1})$.
}
\begin{align}
 \Del{A} &= \cV^{-1}\,\Del{\J}, 
\label{variation1}
\\
 \Del{\bar A} &=\cV^{-1}\,(\ad\, \bar\del)^{-1}\,
  \Bigl( - \cI + \Del{\omega} \Bigr), 
\label{variation3}
\end{align}
where $\delta\omega(\bx)\equiv -\delta \V(\bx)\,\V^{-1}(\bx)$, 
and $\cV(\bx)=\bigl(\cV^{ab}(\bx)\bigr)$ is the matrix representing $\Ad\, V(\bx)$; 
$\V(\bx)\,t^b\,\V^{-1}(\bx) \equiv t^a\,\cV^{ab}(\bx)$.

\noindent
[proof]\\
In general, when a set of fields $\{\Phi_I(\bx)\}$ 
are (locally) transformed to another set $\{\phi_I(\bx)\}$ as
\begin{align}
 \delta \Phi_I(\bx) = \cD_{IJ}\,\delta \phi_J(\bx),
  \quad
  \cD=(\cD_{IJ})=\sum_n d_{i_1\cdots i_n}(\bx)\,\del_{i_1}\cdots \del_{i_n}, 
\end{align}
their variations are transformed as%
\footnote{
This can be proved by noting that the following holds 
for an arbitrary functional $G[\Phi]$: 
\begin{align}
 \delta G[\Phi] 
  &= \int_\bx \delta \Phi_I(\bx)\,\frac{\delta G}{\delta \Phi_I(\bx)}
  =\int_\bx \bigl(\cD_{IJ}\,\delta\phi_J(\bx)\bigr)\,
    \frac{\delta G}{\delta \Phi_I(\bx)}
  =\int_\bx \delta \phi_J(\bx)\,
    \Bigl((\cD^\dag)_{JI}\frac{\delta G}{\delta \Phi_I(x)}\Bigr)
  \equiv \int_\bx \delta \phi_J(\bx)\,\frac{\delta G}{\delta \phi_J(\bx)}.
\nn
\end{align}
}
\begin{align}
 \Del{\phi_I(\bx)}&=(\cD^\dag)_{IJ}\,\Del{\Phi_J(\bx)},
\label{variation4}
\\
 \Del{\Phi_I(\bx)}&=\int_\by (\cD^\dag{}^{-1})_{IJ}(\bx,\by)\,\Del{\phi_J(\by)},
\label{variation5}
\end{align}
where 
$\cD^\dag \equiv \sum_n (-1)^n\del_{i_1}\cdots\del_{i_n}\,
d^{\rm \,T}_{i_1\cdots i_n}(\bx)$. 
We apply this formula to the sets $\{\Phi_I\}=\{A^a,\bar A^a\}$ 
and $\{\phi_I\}=\{\J^a,\omega^a\}$ 
with $\delta \omega^a=-(\delta V\,V^{-1})^a$. 
Equations \eq{variation0a} and \eq{variation0b} 
give 
\begin{align}
 \cD=(\cD_{IJ})=
 \left(
  \begin{array}{ccc}
   \cV^{-1} & 0 \\
    0 & \cV^{-1} \\    
  \end{array}
 \right)
 \left(
  \begin{array}{ccc}
    1 & -\ad\, D\\
    0 & -\ad\, \bar\del 
  \end{array}
 \right),
\end{align} 
so that we obtain
\begin{align}
 \cD^\dag&=
 \left(
  \begin{array}{ccc}
    1 & 0 \\
    \ad\,D & \ad\, \bar\del 
  \end{array}
 \right)
 \left(
  \begin{array}{ccc}
   \cV & 0 \\
    0 & \cV \\
  \end{array}
 \right),
\\
 \cD^\dag{}^{-1}&=
 \left(
  \begin{array}{ccc}
   \cV^{-1} & 0 \\
    0 & \cV^{-1} \\
  \end{array}
 \right)
 \left(
  \begin{array}{ccc}
    1 & 0 \\
    -(\ad\,\bar\del)^{-1}\ad\,D & (\ad\,\bar\del)^{-1}
  \end{array}
 \right).
\end{align} 
Here we have used that $\cV$ is orthogonal, $(\cV^{\rm T})^{-1}=\cV$, 
and $\ad\, D$ and $\ad\, \bar\del$ are anti-self-adjoint, 
$(\ad\, D)^\dag=-\ad\, D$ and $(\ad\, \bar\del)^\dag=-\ad\, \bar\del$. 
Substituting these into \eq{variation5} 
and noting that
\begin{align}
 \ad D_z &= \cV^{-1}\,\ad D\,\,\cV, \quad
  \ad D_{\bar z} = \cV^{-1}\,\bar\del \,\,\cV ,
\end{align}
we obtain \eq{variation1} and \eq{variation3}. [Q.E.D.]

Note that the original Gauss-law operator that generates gauge transformations,
\begin{align}
 \hat{\cI}^a(\bx) \equiv (\ad D_z)^{ab} \Del{A^b(\bx)}
  + (\ad D_{\bar z})^{ab}\Del{\bar A^b(\bx)}, 
\end{align}
can be rewritten with new variables as
\begin{align}
 \hat{\cI}^a(\bx)= \bigl(\cV^{-1}(\bx)\bigr)^{ab}\Del{\omega^b(\bx)} 
 \quad \bigl(\delta\omega = -\delta V\,V^{-1}\bigr). 
\end{align}
Thus, a functional is gauge-invariant 
when and only when it does not depend on $V$.
Note also that \eq{variation3} implies that 
\begin{align}
 \cI = -\ad\,\bar\del\,\,\cV\,\Del{\bar A}+ \Del{\omega}
  = -\ad\,\bar\del\,\,\cV\,\Del{\bar A}+\cV\,\hat\cI.
\end{align}
Thus, for a gauge-invariant functional 
$P[\J]=\hat P[A,\bar A]$ (with $\hat\cI \hat P=0$), 
we have
\begin{align}
 \cI\,P=-\bar\del\Bigl(\cV\,\frac{\delta \hat P}{\delta \bar A}\Bigr). 
\end{align}
This is the formula \eq{hol_Gauss_law2} used in text.

We now investigate properties of the operator defined in \eq{N-bar}, 
\begin{align}
 \bar \cN\equiv -\int(\ad D)^{-1}\bar\del \J\cdot \cV\Del{\bar A}
  =-\int_{\bx,\by}\bigl((\ad D)^{-1}\bigr)^{ab}(\bx,\by)\,\bar\del \J^b(\by)
    \, \cV^{ac}\Del{\bar A^c(\bx)},
\end{align}
when it acts on a holomorphic invariant functional
$P[\J]=P[A,\bar A]$. 
In general, $P[\J]$ can be expressed 
as a formal Laurent series of $\bar \del$ and $D$.
Since $\bar \cN$ is a first-order functional derivative, it obeys 
the chain rule as well as the linearity,
and thus we only need to know the behavior of $\bar\cN$
when acting on a polynomial of $\bar \del$ and $D$ of the same order, 
like
\begin{align}
 P[\J]=\int \bar\del \J \cdot
   (\ad D)^n\,\bar\del \J \quad(n\in\bZ).
\label{2example}
\end{align}
In this example, the orders in $\bar \del$ and $D$ 
are $2$ and $n+2$, respectively 
(recall that $\bar\del \J=[\bar\del,D]$).

We prove that $\bar\cN$ counts the order in $\bar \del$ of
each term in the Laurent series.
We first note that when rewriting $P[\J]$ as a functional of the
original fields $A$ and $\bar A$ (i.e.\ $P[\J]=\hat P[A,\bar A]$),
$\bar \del $ in $P[\J]$ is replaced 
by $\ad D_{\bar z}=\ad(\bar\del+\bar A)$ in $\hat P[A,\bar A]$. 
This is the only possible way for $\bar A$ to appear 
in a gauge-invariant way. 
For example, the polynomial \eq{2example} is rewritten as
\begin{align}
 \hat P[A,\bar A]= \int [ D_{\bar z},D_z]\cdot (\ad D_z)^n
  [ D_{\bar z},D_z].
\end{align}
Due to the fact that the operator $\bar\cN$ is 
a first-order functional derivative with respect to $\bar A$ 
and also that it obeys the chain rule, 
the behavior of the operator $\bar\cN$  
acting on a functional of $P[\J]=\hat P[A,\bar A]$ can be known 
once we understand how the operator acts on a functional 
of simpler form 
\begin{align}
 \hat P'[A,\bar A]&\equiv\int \hat X\cdot [\bar\del+\bar A,\hat Y]
\nn\\
  \Bigl(&=P'[\J]=\int_\bx X\cdot \,\bar\del Y\quad
   \mbox{with $\hat X=\cV^{-1} X$ and $\hat Y=\cV^{-1} Y$}\Bigr) 
\label{nbar1}
\end{align} 
with $\hat X$ and $\hat Y$ being independent of $\bar A$. 
We emphasize again that this simplified functional is 
a representative of the part on which the variation $\delta/\delta \bar A$ acts. 
It is thus sufficient to show that $\bar\cN P' = + P'$, 
which will be proved as follows.

We first notice that 
\begin{align}
 \bar\cN P'&=-\int(\ad D)^{-1}\bar\del \J\cdot \cV\Del{\bar A}
  \int_\bx \hat X\cdot \,\bigl[\bar\del+\bar A, \hat Y\bigr]
\nn\\
 &=-\int \cV^{-1}\,(\ad D)^{-1}\bar\del \J\cdot \Del{\bar A}
  \int_\bx \hat X\cdot \,\bigl[\bar\del+\bar A, \hat Y\bigr]
\nn\\
 &=+\int (\ad D_{z})^{-1}\,\bigl[D_z, D_{\bar z}\bigr]\cdot
  \bigl[\hat Y, \hat X\bigr].
\label{N-bar2}
\end{align}
One could evaluate this by formally deforming the last expression as
$
 \int (\ad D_{z})^{-1}\,\ad D_z\,(D_{\bar z})\cdot
  \bigl[\hat Y, \hat X\bigr] = \int D_{\bar z}\cdot\bigl[\hat Y, \hat X\bigr]
 = \int \bigl[D_{\bar z}, \hat Y\bigr]\cdot\hat X
 =P'.
$
In order to justify such formal manipulations, 
especially to give grounds to the second expression, 
we make a few preparations in the following. 
First we extend the Lie algebra ${\rm su}(N)$ to ${\rm u}(N)$ 
and introduce the set of ${\rm u}(N)$-valued differential operators: 
\begin{align}
 \cS \equiv \bigl\{ \mbox{${\rm u}(N)$-valued differential operator} \bigr\}.
\end{align}
An element $\cD \in \cS$ generically has the following form: 
\begin{align}
 \cD = \frac{1}{i}\,t^\alpha \cD^\alpha
  \quad \Bigl(\alpha=0,1,\cdots,N^2-1;~~
  t^0\equiv \frac{1}{\sqrt{2N}}\bunit_N\Bigr),
\end{align}
where $\cD^a$ is a differential operator. 
For example, the covariant derivatives $D_z$ and $D_{\bar z}$ 
belong to $\cS$ and have the form 
\begin{align}
 D_z=\bunit_N \,\del + \frac{1}{i}t^a A^a,\quad
  D_{\bar z}=\bunit_N \,\bar\del + \frac{1}{i}t^a \bar A^a
  \quad(a=1,\cdots,N^2-1),
\end{align}
where $\del$ and $A^a$ are differential operators with components 
$\del=\bigl(\bra{\bx}\del\ket{\by}=\del\,\delta^2(\bx-\by)\bigr)$ 
and $A^a=\bigl(\bra{\bx}A^a\ket{\by}=A^a(\bx)\delta^2(\bx-\by)\bigr)$, 
respectively. 
We introduce the trace for $\cS$ as
\begin{align}
 \Tr \cD \equiv \int_\bx \tr \frac{\bra{\bx}\cD\ket{\bx}}{\delta^2(\bzero)} 
  \quad (\cD\in\cS), 
\end{align}
which satisfies the relation $\Tr \cD_1\cD_2=\Tr \cD_2\cD_1$ 
for $\cD_1,\,\cD_2\in\cS$. 
The action $\ad \cD$ for $\cD\in\cS$ is then regarded as 
belonging to ${\rm End}(\cS)$, 
and similarly $(\ad \cD)^{-1}\in {\rm End}(\cS)$ 
when its inverse exists in any formal sense 
(which we always assume in the presence of a proper IR regulator).

Now \eq{N-bar2} can be interpreted as taking a trace in $\cS$ 
and can be rewritten in the following way:
\begin{align}
 \bar\cN P'&=-2\int \tr\, \bigl((\ad D_{z})^{-1}\,\bigl[D_z, D_{\bar z}\bigr]\bigr)\,
  \bigl[\hat Y, \hat X\bigr]
  =-2\,\Tr\,\bigl( (\ad D_z)^{-1}\,\ad D_z\,(D_{\bar z})\bigr)\,
  \bigl[\hat Y, \hat X\bigr]
\nn\\
 &=-2\,\Tr\,D_{\bar z}\,\bigl[\hat Y,\hat X\bigr]
  =-2\,\Tr\,\hat X\,\bigl[ D_{\bar z},\hat Y\bigr]
  =\int \hat X\cdot \bigl[D_{\bar z}, \hat Y\bigr]
  =\int X\cdot \bar\del Y
\nn\\
 &=P',\label{nbar2}
\end{align}
which completes the proof of our claim.

Similarly, we can prove that 
$\cN = \int{\bar{\del }}^{-1}\bar{\del }\J\cdot {\delta}/{\delta \J}$
counts the order in $D$ of each term in the Laurent series.
It is again sufficient to show that $\cN P''=+P''$ for a functional
\begin{align}
P''[\J]\equiv\int_\bx X\cdot \bigl[D, Y\bigr] =  \int_\bx X\cdot \bigl[\del + \J, Y\bigr]
\quad\bigl(\mbox{$(X,Y)$: $\J$-independent}\bigr).
\end{align}
This can be performed as follows:
\begin{align}
\cN P''&=\int{\bar{\del }}^{-1}\bar{\del }\J\cdot \frac{\delta}{\delta \J}
      \int_\bx X\cdot \bigl[\del + \J, Y\bigr]\nn\\
     &=\int{\bar{\del }}^{-1}\bigl[\bar \del ,D\bigr]\cdot \bigl[Y, X\bigr] \nn\\
     &=-2\,\Tr\,\bigl( (\ad \bar \del)^{-1}\,\ad \bar \del \,(D)\bigr)\,
  \bigl[Y, X\bigr]\nn\\
     &=-2\,\Tr\,D\,\bigl[Y,X\bigr]
     =-2\,\Tr\,X\,\bigl[ D,Y\bigr]\nn\\
     &=P'' ~.
\end{align}


\setlength{\itemsep}{5.\baselineskip}

\end{document}